\documentclass[namedreferences]{solarphysics}

\usepackage[hyperref,optionalrh,showbiblabels]{spr-sola-addons} 
\usepackage{graphicx}        
\usepackage{color}           
\usepackage{breakurl}        

\chardef\us=`\_

\begin{document}

\begin{article}
\begin{opening}

\title{Quasi-periodic pulsations in solar and stellar flares: an overview of recent results}

\author[addressref=aff1,corref,email={tom.vandoorsselaere@wis.kuleuven.be}]{\inits{T.}\fnm{Tom}~\lnm{Van Doorsselaere}\orcid{0000-0001-9628-4113}}
\author[addressref={aff1,aff2}, email={elenku@bk.ru}]{\inits{E.G.}\fnm{Elena G.}~\lnm{Kupriyanova}}
\author[addressref=aff3, email={DYuan2@uclan.ac.uk}]{\inits{D.}\fnm{Ding}~\lnm{Yuan}}

\address[id=aff1]{Centre for mathematical Plasma Astrophysics, Department of Mathematics, KU Leuven, Celestijnenlaan 200B bus 2400, 3001 Leuven, Belgium}
\address[id=aff2]{Central Astronomical Observatory at Pulkovo of the Russian Academy of Sciences, 196140, St. Petersburg, Russia}
\address[id=aff3]{Jeremiah Horrocks Institute, University of Central Lancashire, Preston, PR1 2HE, United Kingdom }

\runningauthor{Van Doorsselaere et al.}
\runningtitle{QPP in solar and stellar flares}

\begin{abstract}
Quasi-periodic pulsations (or QPPs) are periodic intensity variations in the flare emission, across all wavelength bands. In this paper, we review the observational and modelling achievements since the previous review on this topic by \citet{nakariakov2009qpp}. In recent years, it has become clear that QPPs are an inherent feature of solar flares, because almost all flares exhibit QPPs. Moreover, it is now firmly established that QPPs often show multiple periods. We also review possible mechanisms for generating QPPs. Up to now, it has not been possible to conclusively identify the triggering mechanism or cause of QPPs. The lack of this identification currently hampers possible seismological inferences of flare plasma parameters. 
QPPs in stellar flares have been detected for a long time, and the high quality data of the Kepler mission allows to study the QPP more systematically. However, it has not been conclusively shown whether the time scales of stellar QPPs are different or the same as those in solar flares.
\end{abstract}
\end{opening}

\section{Introduction}
     \label{S-Introduction} 
     
In the past 15 years, it has become obvious that the solar corona hosts a variety of magnetohydrodynamic (MHD) waves. First, slow waves were detected in coronal plumes and loops \citep{deforest1998, berghmans1999}, quickly followed by kink waves excited by flares \citep{nakariakov1999}. For an overview of the earlier observations, we recommend a thorough reading of \citet{nakariakov2005} or \citet{demoortel2012b}. More recently, it was found that kink waves are observed in nearly all coronal loops \citep{tomczyk2007,vd2008,mcintosh2011,nistico2013}, and that slow waves are also detected in all coronal loops at all times \citep{krishnaprasad2012,banerjee2015}. \par
\begin{figure}
	\centerline{\includegraphics[width=.5\linewidth]{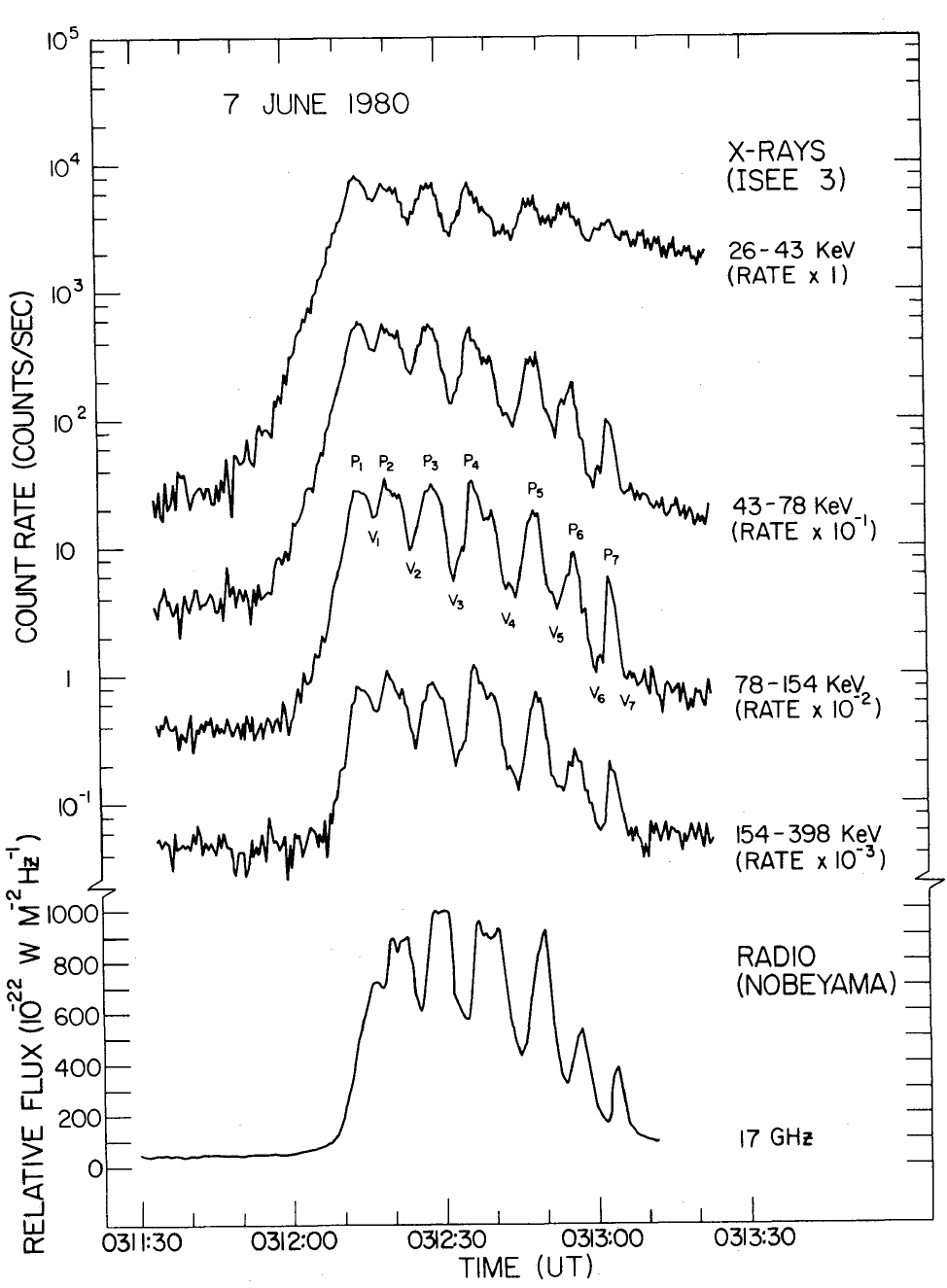}}
	\caption{Intensities as a function of time for the ``7 sisters'' flare. Picture taken from \citet{kane1983}.}
	\label{fig:7sisters}
\end{figure}

The violent nature of the flare is bound to excite waves in its vicinity. Often, the direct observations of flares and their surroundings show clear oscillatory phenomena. Indeed, it is obvious that transverse waves are excited in nearby loops by the flare disturbance. In particular, the flare light curve shows periodic intensity increases and decreases. These are called {\em quasi-periodic pulsations (or QPPs)}. The classical example is the so-called ``7 sisters'' flare \citep{kane1983}, as depicted in Figure~\ref{fig:7sisters}, where a clear 8 s period is observed.
QPPs in solar flares have typical periods between a few seconds to a few minutes. For stellar flares, the periods may reach tens of minutes. These time scales puts the QPP oscillations in the MHD regime. \par
QPPs were reviewed in depth by \citet{nakariakov2009qpp}. The current article aims to give an overview of recent theoretical and modelling results. Even though many new results have been obtained, still there is no consensus reached on what physical mechanism is responsible for the generation of QPPs. \par
The importance of studying QPPs cannot be overstated. If the QPPs are generated by quasi-steady processes (also called magnetic dripping), their period probably contains information on the ongoing physical processes during the flare. If they are generated by magnetic dripping or MHD waves, they offer us a unique tool to seismologically probe the flare site. By comparing the QPP periods to theoretical models, it is possible to gauge plasma parameters in the flare surroundings, which are difficult to measure otherwise (\textit{e.g.} magnetic field). Moreover, it has been argued by \citet{fletcher2008,russell2013} that MHD waves are responsible for the energy transport from the reconnection site to the main emission sites in the flare foot-points. The QPPs could be a direct sign of this transport mechanism.\par
Up to now, however, it is unclear what causes the QPPs (as mentioned above). This hampers progress in using QPPs for seismological purposes (using either the quasi-steady/dripping or wave description).

\section{Observations of QPPs in solar flares}
\label{S-aug}

Since QPPs are typically short-period phenomena \citep[the typical periods of seconds are short compared to other oscillatory phenomena in the solar atmosphere,][]{nakariakov2005}, they are mostly observed with instruments observing the Sun as a star, \textit{i.e.} with no (or limited) spatial resolution. Key instruments are often radio telescopes, because they offer a superior time resolution. Because of the nature, history and observability of the QPPs, we will first focus on the temporal domain (Section~\ref{sec:time}), before moving to the spatial (Section~\ref{sec:spatial}) and spectral (Section~\ref{sec:spectral}) information.

\subsection{Temporal behaviour}
\label{sec:time}

Most studies focus on frequency analysis of the time series of a flare emission intensity. This is usually done by subtracting some flare trend (established by smoothing the flare light curve, Fourier or wavelet filtration, polynomial fitting, or other techniques), before analysing it with standard period analysis techniques (such as FFT, nonlinear fitting, or wavelets). However, recently, some new methods have been exploited \citep{kolotkov2015,inglis2015}. \citet{kolotkov2015} focus on the detection of non-linear and time-dependent signals \citep[\textit{e.g.}][which has a clear anharmonic signal indicating the presence of non-linear waves]{2010PPCF...52l4009N} with the Hilbert-Huang transform implementing the Empirical Mode Decomposition. \citet{inglis2015} focus on the proper accounting of the flare noise. \par
The periods of QPPs in solar flares cover a wide range, but are rather short. They go from sub-second, to seconds, tens of seconds, with maximal periods of around 5 minutes. \citet{tan2010} reported an event in which even all of these periods were simultaneously present in their data. Indeed, this has become abundantly clear in the last 5-6 years: QPPs often show multiple periods in the same flare and are thus multiperiodic phenomena. \citet{inglis2009b} were one of the first to study multiperiodicity in a flare. They found a spectrum in which the periods were nearly multiples of each other. This was unexpected, because non-dispersive waves would rather have frequencies that are multiples of the fundamental mode. Further multiperiodic events were reported by \citet{vd2011b, kupriyanova2013, kupriyanova2013b, kupriyanova2014, kolotkov2015, chowdhury2015}. In some of these cases, the periods seem to be associated with harmonics of a single wave mode \citep[\textit{e.g.}][]{kupriyanova2013,chowdhury2015}. However, in other cases, the periods are far removed from each other \citep{vd2011b}, probably indicating they come from different wave modes. \par
In most cases, the periods seem to be stable. However, in other cases, the periods are observed to grow. For instance, \citet{reznikova2011} found that the QPP period grew from 2.5 to 5 minutes over the duration of the flare, and \citet{huang2014} observed a period growing from 21-23 s in soft X-rays and microwave emission, to 24 s in the EUV and even 27-32 s in radio, with these subsequent bands associated with later or outgoing phases of the flare.  \par
At the moment, it is unclear what the relationship is between the phases of the flare and the timing of the occurrence of the QPP. In some cases, the QPPs occur in both the rise phase and the gradual (decay) phase of the flare \citep[\textit{e.g.}][]{vd2011b,dolla2012,simoes2015}. However, in some cases, the QPPs are restricted to only the rise phase \citep[][and follow-up works]{jakimiec2012}, but in other cases, the QPPs are observed during the decay phase only \citep[\textit{e.g.}][]{kane1983}. In the latter work, very regular pulsations were observed during the decay phase of the flare, lasting for 7 cycles (see Figure~\ref{fig:7sisters}). On the other hand, often short damping times are observed, with only a few cycles visible \citep[\textit{e.g.}][]{kolotkov2015}. Thus, it seems that the quality factor of QPPs has a broad range (quality factor is the damping time divided by the period, \textit{i.e.} some are damped in a few periods, while others are nearly non-damped), indicating that multiple physical mechanisms may be responsible for generating QPPs. The damping time could be heavily influenced by the signal noise and the relative strength of the QPP compared to the flare background.\par
Another point of note is that the QPPs are not co-temporal in different wavelength bands, and that there are time delays observed. \citet{zimovets2010} showed a phase shift for a long period QPP. \citet{dolla2012} performed an extensive study using all available data sources and showed that the phase shifts could reach up to a quarter or even half of the QPP period in some wavelength bands.\par
An important question that may be asked is whether QPPs are a frequently observed phenomenon. This is certainly true, as was proven observationally by \citet{kupriyanova2010}. They selected 12 standard flares (standard in the sense of isolated and close enough to the disk centre), and found that 10 out of 12 showed QPPs. This is convincing evidence that QPPs are an inherent feature of solar flares: solar flares without QPPs are rarely observed. On the other hand, \citet{mossessian2012} applied their technique to 412 bursts and found that only in 2.5\% of them the QPPs appear in the dynamic spectrum, and in 10\% of them the QPPs are present at only a few frequencies. Moreover, it was argued recently by \citet{gruber2011,inglis2015} that the flare variations follow red noise  (\textit{i.e.} a power law in Fourier space, rather than uniform white noise).\par 
Indeed, such power laws are prevalent in solar and astrophysics: it has been observed in the Fourier spectrum of narrowband dm-radio spikes \citep{2005AdSpR..35.1799F} and zebra patterns \citep{2014A&A...561A..34K}. A similar power-law Fourier spectrum was found during a solar flare \citep{2005A&A...432..705K}, in which QPP periods in the range of 0.9--7.5 s were found simultaneously. Also, \citet{liu2011} found indications for the power-law distribution of peaks in the Fourier spectrum of the EUV emission of a C3.2 solar flare/CME event. It seems to be a general feature of coronal emission and waves \citep{2008ApJ...677L.137B,2009ApJ...697.1384T,2014A&A...568A..96G,2015ApJ...798....1I}, and the last believed that this Fourier spectral shape contradicts the commonly held assumption that coronal time series are well described by the sum of a long-timescale background trend, oscillatory signal, and normally distributed noise. The power-law shape of the Fourier power spectrum is interpreted as being due to the summation of a distribution of exponentially decaying emission events along the line of sight. \par
Thus, it is the opinion of \citet{inglis2015} that the QPP identification and detection needs to be revisited with this in mind. They stated that the QPPs should have a power higher than the 99\% confidence level using the red noise approach, \textit{i.e.} in the confidence level above the power law in Fourier space (rather than the confidence above the white noise, as is commonly assumed). This was taken into account by \citet{simoes2015}, who performed a large study of 35 flares. They still found with 97.5\% confidence that most flares (80\% of 35 flares) had significant oscillations, above the red noise level. \par
In the literature, it is clear that no firm concensus has been reached on the occurrence rate of QPPs. \citet{kupriyanova2010,simoes2015} found that QPPs are very common, but \citet{mossessian2012} found that they are rare. It should be investigated if there are selection effects in these articles for selecting the events, or for the period detection.

\subsection{Spatial location}\label{sec:spatial}
\begin{figure}
	\centerline{\includegraphics[width=.6\linewidth]{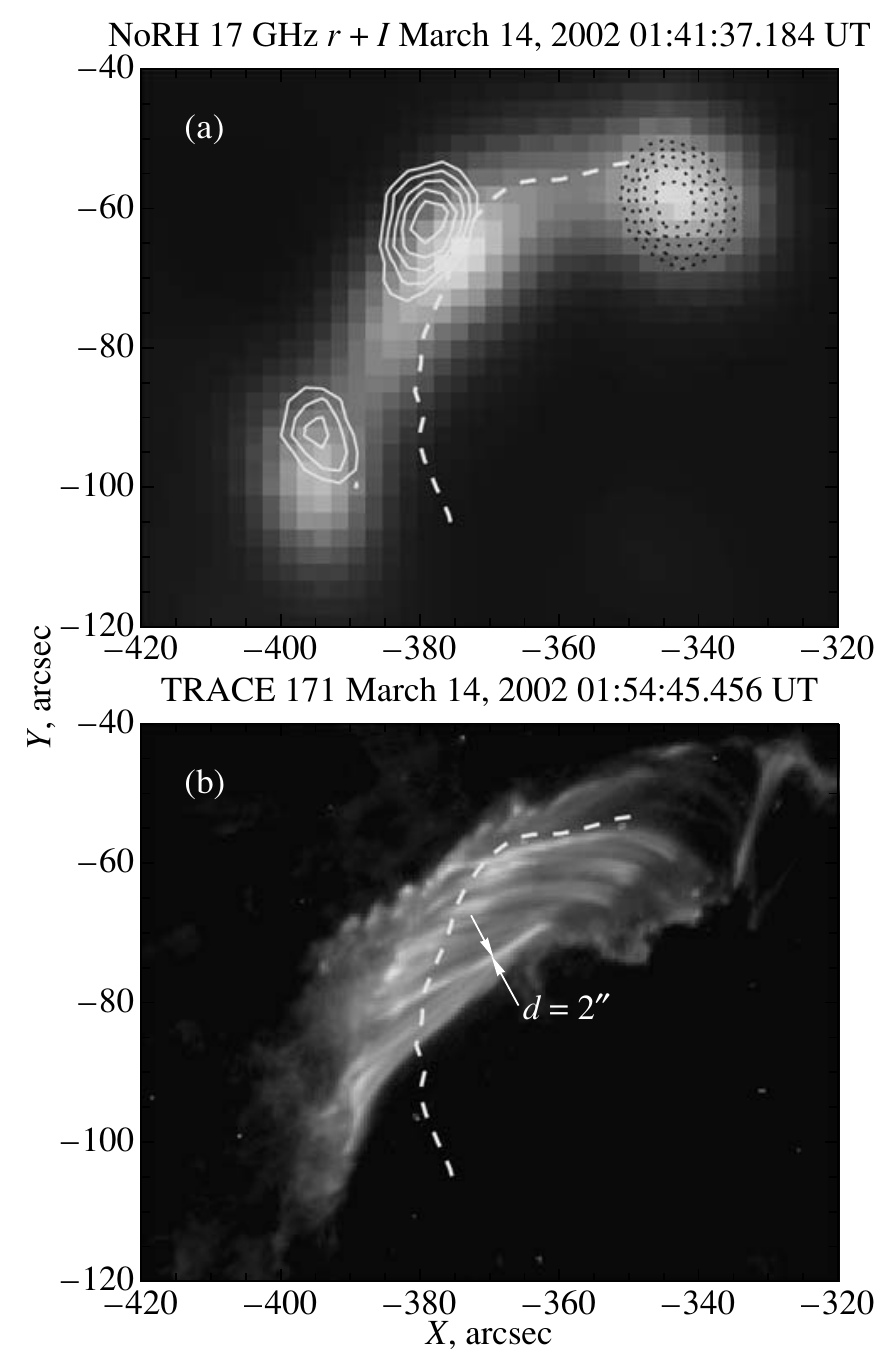}}
	\caption{A re-analysis of a QPP observed in the \textit{Nobeyama RadioHeliograph} (top) with the TRACE instrument (bottom), showing that a single flare loop in radio may correspond to a bundle of flare loops in EUV. Figure taken from \citet{zimovets2013}.}
	\label{fig:zimovets}
\end{figure}

Aside from the temporal evolution of the flare emission, there is currently a trend to investigate spatially resolved flares. This was already possible with the most modern radio telescopes, such as the \textit{Nobeyama RadioHeliograph}. For example, \citet{melnikov2005} could identify the fundamental sausage mode and its overtone in a coarsely resolved loop. However, the radio observations often have a coarse spatial resolution. This was stressed when \citet{zimovets2013} re-analysed a seemingly resolved flare observed by \citet{kupriyanova2010}. \citet{zimovets2013} pointed out that the single loop flare in the radio band actually corresponded to a bundle of unresolved flaring loops in TRACE (see Figure~\ref{fig:zimovets}). They argued that it was the propagation from one loop to the next that created the QPPs in this case, rather than a sausage mode in a thick flaring loop.\par
It is well known that solar flares excite transverse waves in nearby loops \citep[see, \textit{e.g.}][]{2013SoPh..282..523K, nistico2013, 2013ApJ...774..104W, 2015A&A...577A...4Z,goddard2016}, and sometimes also loop contractions \citep[see, \textit{e.g.}][]{2012ApJ...757..150L, 2013RAA....13..526Z}.  In recent observations of \citet{simoes2013}, it seemed that the transverse oscillations of contracting loops were going hand in hand with QPPs. They found that the transverse oscillations of surrounding loops were apparently caused by the ``persistent, semi-regular compression of the flare core region''. This is an observational indication of the nature of the QPP, which is interacting directly with (some of) the surrounding loops. This effect \citep[as also suggested by][]{2015A&A...577A...4Z} was observed and modelled by \citet{2015A&A...581A...8R} for five contracting loops during a solar flare. The authors found that the highest loops oscillated during their contraction while the shortest loops did not. They concluded that coronal loop contractions and oscillations together can be caused by removal of magnetic energy from the corona, which could be related to the QPP. The proposed framework suggests that loop motions can be used as a diagnostic for the removal of coronal magnetic energy by flares \citep[confirming the observational results of][]{simoes2013}, and the rapid decrease of coronal magnetic energy is a newly-identified excitation mechanism for transverse loop oscillations \citep{2008ASPC..383..221H}.\par
Evidence for an external trigger of the QPPs (Section~\ref{sec:external}), or a common trigger for external oscillations, was found in EUV observations. It was found by \citet{liu2011} that the QPP periods agreed with quasi-periodic fast propagating (QFP) waves in a nearby funnel, indicating a common origin. Another possible connection with external waves was revealed from AIA observations of slow waves. \citet{kumar2015} found long period slow waves propagating in a flaring loop, with a period of 409 s. Simultaneously, a QPP with a period of 202 s was observed, of which the emission originated in one foot-point. The physical connection between these two periodicities is unclear, but it could be related to a resonant coupling between the QPP and the slow wave by the 2nd harmonic of a wave mode. The quasi-periodic response at another foot-point suggests a closed configuration of the magnetic field lines. In an open magnetic field configuration, the periodicity appears as quasi-repetitive type III radio bursts \citep{2016ApJ...822....7K} produced by ultra relativistic electrons escaping the solar coronae. The period of about 3 minutes corresponds to the periods of the sunspot's oscillations, and some observational evidence has been found for a link \citep{sych2009}.
\par
In \citet{2012ApJ...754...43S} and \citet{2012ApJ...755..113S}, the formation of cusp-like structures from a loop system is seen directly in imaging observations with the SDO/AIA facilities. \citet{2012ApJ...754...43S} detected bidirectionally propagating waves in the 171\ \AA{} in one of the loop legs. The waves corresponding to the QPPs were found to be directed upward only, above the cusp. The intensity modulation in all wavelengths was large, more than 30\% of the emission trend for the event. The authors estimated that the magnitude exceeded that expected for density perturbations caused by slow-mode magnetoacoustic waves. So, the authors connected their observation of the QPP with episodic outflows along the tube, caused by the physics of the reconnection (Section~\ref{sec:flare}).\par
A similar QPP event was studied by \citet{2012ApJ...755..113S}. QPPs were detected in almost all the loop sources. The location of the QPPs was different for separate phases of the flare. Before the flare, pulsations were seen in one loop only. After the flare, pulsations were pronounced in the newly appeared cusp and loop. The pulsations in the new structures were 5 to 7~minutes delayed relatively to the ones in the old loop. However, they were found to correlate well after removal of the delay. The QPP periods range from 24 to 166s. The span of the periods found with the time-distance diagrams in different parts of the loop system were explained by the presence of different MHD modes (see Section~\ref{sec:postflare}). \par
Observational data in the microwave and ultraviolet bandpasses reveal that oscillatory processes in the sunspot led to the flare ignition in a neighbouring active region \citep{2015A&A...577A..43S}.  Spatial analysis of the sunspot oscillations revealed that a magnetic waveguide connected the sunspot with the flare site. The waveguide was also well pronounced in the coronal line emission (SDO/AIA 171\ \AA). Slow magnetoacoustic waves propagated from the sunspot to the flare region along this magnetic channel, and this argues again in favour of an external triggering mechanism (Section~\ref{sec:external}). 

\subsection{Spectral information}
\label{sec:spectral}
What is lacking mainly in the current observational QPP studies is the spectral information. There is a wide coverage of broadband imaging facilities, ranging from $\gamma$-rays \citep{nakariakov2010} to radio emission. However, detailed spectroscopic studies of QPPs have not been performed. \par
In general for flares, a lot of progress has been made by \citet{milligan2012,milligan2012b,milligan2013}. They used the MEGS spectrometer in the EVE instrument on board SDO to show that continuum emission plays a major role in the flare emission. Thus, the question is raised how QPPs would be observed in the spectral domain. Are they mainly observed as variations of the continuum (as seems to follow from \citeauthor{milligan2013}), or would they show up in varying intensity of hot spectral lines? In the stellar flares studies, the spectral approach is often taken \citep[\textit{e.g.}][]{kowalski2013} with great success, and this urges the same approach in QPP observations.\par
An interesting observational finding has been made by \citet{2015ApJ...807...72L}. They performed a multi-wavelength study of a solar flare. They showed that there are many bright structures (in AIA 1600\ \AA{} images) moving along the flare ribbon parallel to the magnetic neutral line. The periodicity of the QPP of 4 minutes are found also in X-rays, EUV lines, radio emission, and Doppler velocities. From the imaging data in the EUV range, it was concluded that the QPPs originate from the flare ribbon front. Spectral observations with IRIS indicate that both the spectral line width and Doppler velocity were varying in phase with the QPP.\par
In a very recent study, \citet{brosius2015} analysed IRIS observations of solar flare ribbons. In IRIS, the Fe {\small XXI} spectral line together with cooler chromospheric lines (such as O {\small IV} and C {\small IV}) are observed simultaneously in the same wavelength band. \citeauthor{brosius2015} detected QPPs in the emission of the cooler lines, and that they were not as clear in the hot Fe {\small XXI} lines. Moreover, they also observed simultaneous Doppler shift oscillations in the cooler lines. The Doppler shift oscillations are in phase with the intensity oscillations, similar to \citet{2015ApJ...807...72L}. Thinking ahead to a possible interpretation in terms of flare loop wave modes (Section~\ref{sec:postflare}), this places strong restrictions on the possible wave mode identification.

\section{Interpretations for QPPs}
The periodic behaviour of QPPs must have a physical explanation and interpretation. Sometimes it is said that multiple flares (or sympathetic flares) occur sufficiently close in time, to show up as a secondary peak in the light curve. However, this does not explain the occurrence of nicely periodic intensity variations, because the chance of having multiple flares occurring periodically is rather small, except in a few idealised situations or models.\par 
Given the periodic nature, it is straightforward to study oscillatory plasma processes to explain the QPPs. The typical time scale of a few seconds to a few minutes is firmly in the MHD regime, and thus this is what we will be focussing on in the explanations. 

\subsection{Physical mechanism}\label{sec:models}
\begin{figure}
	\centerline{\includegraphics[width=.6\linewidth]{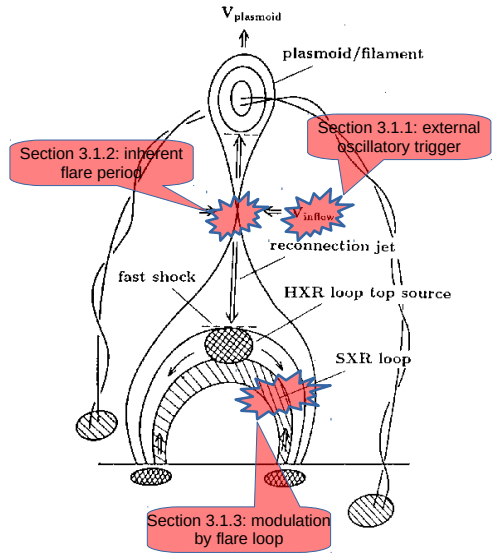}}
	\caption{Schematic representation of possible causes for the generation of QPPs. The QPPs could be caused in the external flare driver, in the flare itself or in the post-flare loops. Figure adapted from \citet{shibata1995}.}
	\label{fig:qpps}
\end{figure}

It is currently debated what physical mechanism is responsible for the generation of QPPs. Three possible mechanisms will be discussed in detail: waves could be excited by the flare trigger (Section~\ref{sec:external}), in the flare itself (Section~\ref{sec:flare}), or in the post-flare loops (Section~\ref{sec:postflare}), as shown schematically in Figure~\ref{fig:qpps}. 
For each proposed mechanism, we will list the observational arguments that support or contradict that particular theory. 

\subsubsection{External oscillatory triggers}\label{sec:external}
\begin{figure}
	\centerline{\includegraphics[width=.6\linewidth]{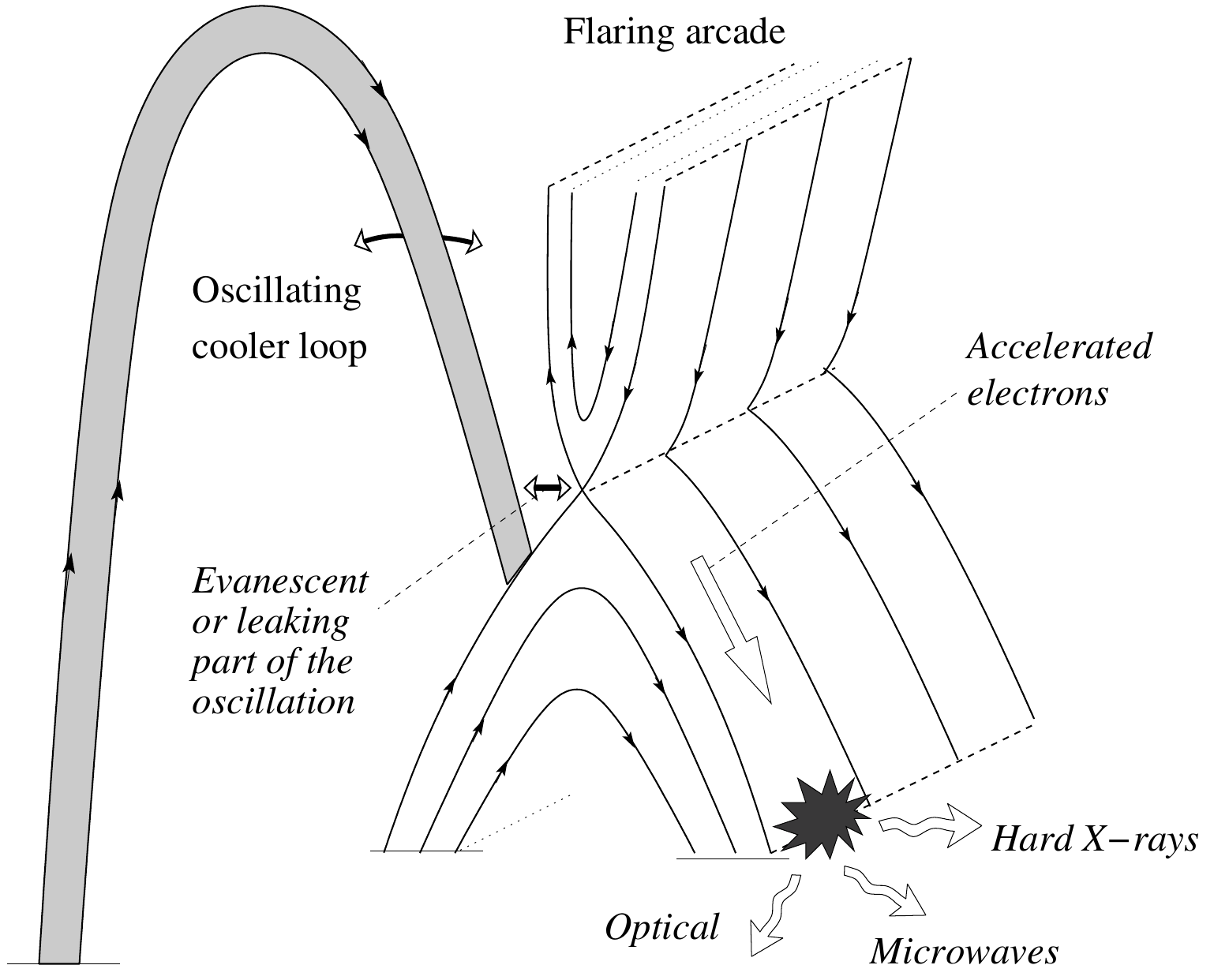}}
	\caption{Schematic drawing of an external periodic triggering of reconnection, resulting in periodic precipitation of energetic particles and modulation of the observed emission. Figure taken from \citet{nakariakov2006}.}
	\label{fig:external}
\end{figure}

The first possibility is that QPPs are triggered by a periodicity in the inflow that drives the reconnection. Then, the reconnection is modulated periodically, leading to a periodic production of non-thermal particles. This then naturally explains the modulation in the emission, by the episodically increased and decreased number density of non-thermal particles precipitating in the flare loop foot-points. This possibility for generating QPPs was first suggested in \citet{nakariakov2006}, and is shown schematically in Figure~\ref{fig:external}. \par
In the figure, the indicated cooler loop symbolises the external trigger for the QPPs. The external trigger is an essential ingredient in this scenario. As shown in the picture, it could be a transversely oscillating loop, for which the evanescent tail extends to the null-point. It could also be a propagating disturbance generated by \textit{e.g.} another flare. Possible triggers may be any oscillatory disturbance reaching (the surroundings of) the null-point and thereby periodically triggering/modulating the reconnection rate. \par
The first models for oscillations were developed by \citet{mclaughlin2004}, where they showed that a wave pulse steepened around the null-point, leading to the generation of strong currents (and thus reconnection). Later on, \citet{nakariakov2006} extended the experiment to mimic periodic external drivers, pointing out that the small-amplitude external driver results in periodic current spikes near the null point, of which the amplitude is much larger than the amplitude of the driving wave, because of the non-linear behaviour near the null point.\par
Further study on this subject was performed by \citet{gruszecki2011} and they incorporated non-linear wave effects (such as shock dissipation in the models). The shock dissipation in the incoming wave will damp the incoming waves with high amplitude, while the low-amplitude waves will not have a strong effect. Thus, the external triggering wave needs to have the amplitude ``just right'' to have maximal effect on the  null point (\textit{i.e.} a Goldilocks principle). This could explain the difference in strength between QPPs in different flare events, because they strongly depend on the amplitude of the external driver.\par 
More numerical simulations were done for a more general magnetic field configuration by \citet{calvo2015}. They showed that a low frequency driver (with periods around 5 minutes, typical in the solar atmosphere) can lead to high frequency oscillations ($>10\ \mbox{mHz}$) with periods close to those of QPPs. It is unclear for now how the short-period waves are generated from the longer-period waves, but it could be related to dispersive evolution or resonant filtering. \par
Even though their model is entirely different, the work by \citet{zimovets2011} can also be sorted under this category (``external oscillatory triggers''). They proposed the cross-field propagation of slow waves to explain the apparent propagation of the flare foot-points. This could also lead to a periodic triggering of the reconnection along an X-line resulting in QPPs. However, with this mechanism the time scales may be too slow to explain the short QPP periods. As similar (albeit more convoluted) model was described in \citet{artemyev2012}, who focussed on waves above post-flare arcades. \par
Some observational support for the scenario in which QPPs are triggered by external oscillations is offered by the work of \citet{liu2011}. They found that the light curves from the flaring region shows the same periodicity as for the quasi-periodic fast magnetoacoustic propagating wave train in the nearby funnel of open magnetic field. This suggests that both phenomena have a common cause, externally linked to the flare region. \par
The scenario of externally triggered QPPs can also clearly explain the time delays between QPPs in different wavelength bands \citep{dolla2012}. In this scenario, we would have the co-temporal, periodic generation of the non-thermal particles near the X-point. Due to the different propagation speeds of the species and their different dissipation location (where the non-thermal particle is thermalised and emitting), this would naturally lead to time delays between wave signals in different wavelength bands.\par
Still, it is hard to explain multiperiodic QPPs with this scenario, because it is unclear how multiple external periods could be propagated through the highly non-linear reconnection response. Further modelling is needed to see a multiperiodic driver could result in multiperiodic QPPs and how the X-point filters or amplifies different harmonics in the external driver or possibly generates them itself. \par
As mentioned in Section~\ref{sec:spatial}, some observational evidence exists from spatially resolved QPP events. In \citet{liu2011} there seems to be a connection with the quasi-periodic fast propagating (QFP) waves, and in \citet{kumar2015} there is evidence for an apparent harmonic connection between the QPP and the propagating slow wave.

\subsubsection{Periods generated by the flare}\label{sec:flare}
A second possible generation mechanism for QPPs is the built-in periodicity of the reconnection in the flare. Even though the inflow into the reconnection region may be non-periodic, the reconnection is prone to generate waves and is inherently an unstable process.\par
\begin{figure}
	\centerline{\includegraphics[width=\linewidth]{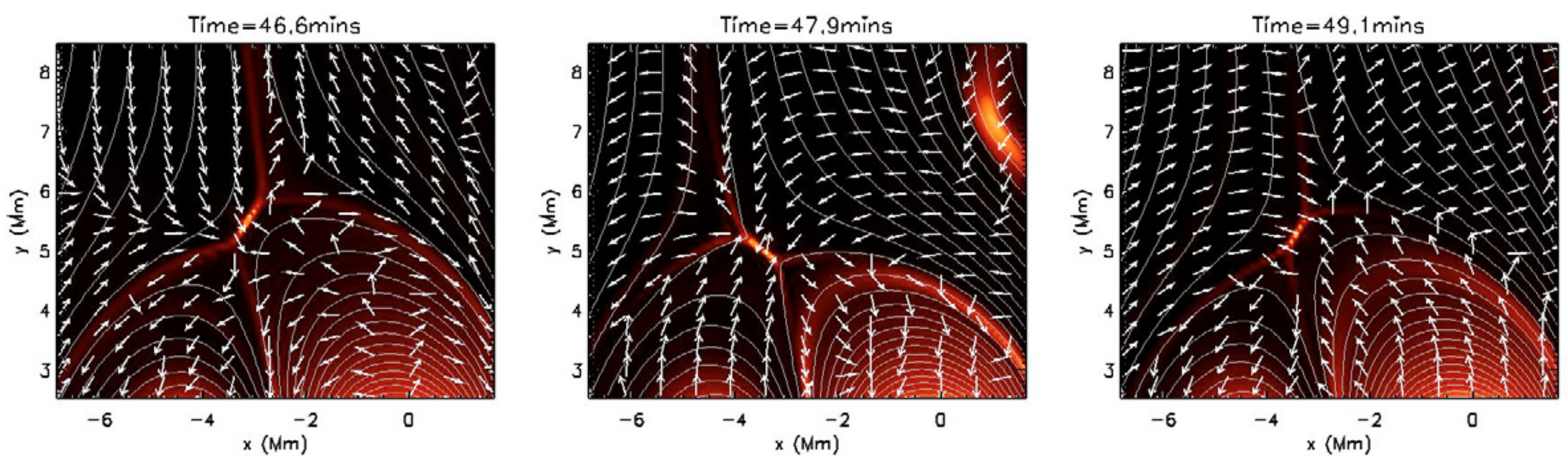}}
	\caption{Figure taken from \citet{murray2009}, showing the temporal evolution of the current in their numerical model.}
	\label{fig:craig}
\end{figure}

In a solar physics context, flare oscillations were first studied by \citet{craig1991} and follow-up works. They considered a resistive X-point surrounded by a fixed, circular boundary and studied the eigenoscillations of the system. They found that there is a damped oscillatory regime of the reconnection, in which the angle between the X-point field lines is changed periodically (see also representative simulations in Figure~\ref{fig:craig}). Thus, these waves are fast waves, of which the period is determined by the Alfv\'en speed profile around the X-point, which forms the resonator. The damping of the wave is governed by the resistivity of the plasma. \par
The oscillatory reconnection was found in 2.5D numerical simulations by \citet{murray2009} (see Figure~\ref{fig:craig}) and in 2D by \citet{mclaughlin2009}. In the first work, an oscillatory regime of reconnection is shown, similar to \citet{craig1991}, and this leads to a periodic variation of the reconnection rate and current density in the current sheet. \citet{mclaughlin2009} study the effect of an incoming wave on the reconnection, and find once again that the current sheet deforms and oscillates. However, in their work, they find that the reconnection jets (and the associated pressure) play a role in the dynamics of the current sheet, and thus the oscillation mode may be determined by the times scales of the slow magnetoacoustic waves transit time, rather than the Alfv\'en transit time. This was later also studied by \citet{mclaughlin2012} in a similar setting, where they performed a parametric study. \citet{mclaughlin2012b} reported that the period of the oscillatory reconnection depends on the amplitude of the exciting wave, indicating that the waves are essentially non-linear. \citet{pucci2014} perturbed an X-point with magnetosonic perturbations, and also found oscillatory reconnection. It was most clearly visible in the out-of-plane vorticity, and this seems to indicate a fast nature of the oscillations.\par 
In any case, the eigenoscillations of the Alfv\'en speed resonator around an X-point may result in the periodic production of energetic particles which modulate the flare emission to a QPP. The (presumably) Sturm-Liouville spectrum of this Alfv\'en speed resonator could also naturally lead to multiperiodic QPPs. This mechanism for the generation of QPPs would be extremely nice for remote sensing with seismology: with the QPP wave signals we could directly probe the reconnection site for the Alfv\'en speed profile. Still, more modelling work is needed to see how the wave periods depend on the current sheet properties and what sort of wave spectrum would be formed. On the other hand, it could be be very difficult to observationally verify this generation mechanism for QPPs.\par
Even if the X-point does not serve as a resonator, it is to be expected that reconnection is a source of waves. This was explored by \citet{longcope2007}. The reconnection and its outflows excite multiple wave modes in the flaring environment. The same idea was used in the work by \citet{takasao2015}. They initiated reconnection above a magnetic arcade in a 2D MHD simulation. They found that the flare outflow led to the formation of fast-mode shocks near the arcade top. The resonance between these two fast-mode shocks then led to periodicity in the flare outflow (which would precipitate to become the QPP emission). It is unclear if the mechanism found by \citet{takasao2015} could lead to multiperiodicity in the emission.\par
From numerical experiments, we also know that the tearing mode instability is operating in the current sheet. Thus, there is a quasi-periodic generation of plasmoids in the current sheet, which get subsequently ejected. Theoretical models for QPPs with this scenario were developed by \citet{tan2007,tan2008}, yielding a broad range of possible QPP periods \citep{tan2010}. Following early numerical simulations \citep{kliem2000}, more advanced simulations concerning solar flares were performed by \citet{barta2008}, but the concept is studied in other fields of plasma physics as well. The QPPs can then be generated by the precipitation of the plasmoids in the foot-points, leading to periodically increased emission. Recent work using particle-in-cell simulations by \citet{innocenti2015} shows that the generation of plasmoids stops after the formation of a monster plasmoid. Therefore, it may be that the plasmoid generation could not explain long-lived QPPs. Moreover, it is unclear how the (chaotic) tearing instability may lead to such cleanly periodic signals in the observations. \par
The coalescence instability of current-carrying loops was previously also considered as a generation mechanism for QPPs, but since no recent work was done on this topic, we refer the reader back to \citet{nakariakov2009qpp} for more details.

\subsubsection{MHD waves in flaring system}\label{sec:postflare}
A third option indicated in Figure~\ref{fig:qpps} is that there are oscillations present or excited in the flaring or post-flare loops. As mentioned before, waves are currently considered as a transport mechanism for the flare energy from the X-point location to the foot-points \citep{fletcher2008,russell2013}. Thus, it is natural to assume that they are reflected several times in the flaring loop, setting up standing waves in the resonant cavity. \par
A flaring loop is often modelled as a radially piecewise constant, static cylinder (with radius $R$) in magnetohydrostatic pressure equilibrium \citep[\textit{e.g.}][]{vd2011b}, following the early work of \citet{edwin1983}. This is questionable, because it is sure that flaring loops have strong flows (\textit{i.e.} not static) and not in equilibrium, let alone pressure equilibrium. However, the model is surprisingly robust, and the basic wave solutions survive when flows \citep{goossens1992} or non-equilibria \citep{ruderman2011} are considered. The model independently constructed by \citet{edwin1983,zaitsev1975} results in a dispersion relation for the frequency $\omega$:
\begin{equation}
	\frac{\kappa_\mathrm{i}}{\rho_\mathrm{i}(\omega^2-k_z^2V^2_\mathrm{Ai})}\frac{J'_m(\kappa_\mathrm{i}R)}{J_m(\kappa_\mathrm{i}R)}=\frac{\kappa_\mathrm{e}}{\rho_\mathrm{e}(\omega^2-k_z^2V^2_\mathrm{Ae})}\frac{K'_m(\kappa_\mathrm{e}R)}{K_m(\kappa_\mathrm{e}R)},
	\label{eq:cylinder}
\end{equation}
where $k_z$ is the longitudinal wave number. The integer number $m$ is the wave number in the azimuthal direction, $\rho$ is the mass density, and $V_\mathrm{A}$ is the Alfv\'en speed. The subscripts $\mathrm{i}/\mathrm{e}$ denote quantities inside/external to the loop. The Bessel function of the first kind is denoted with $J_m$, and the modified Bessel function of the second kind with $K_m$. The prime $'$ denotes a derivative with respect to the argument. $\kappa$ takes the role of radial wave number and is calculated from the wave speeds by
\begin{equation}
	\kappa^2=\left|\frac{(\omega^2-k_z^2 V_\mathrm{S}^2)(\omega^2-k_z^2V_\mathrm{A}^2)}{(V^2_\mathrm{S}+V^2_\mathrm{A})\omega^2-k_z^2V^2_\mathrm{S}V^2_\mathrm{A}}\right|,
\end{equation}
in which $V_\mathrm{S}$ is the sound speed.\par
\begin{figure}
	\centerline{\includegraphics[width=.8\linewidth]{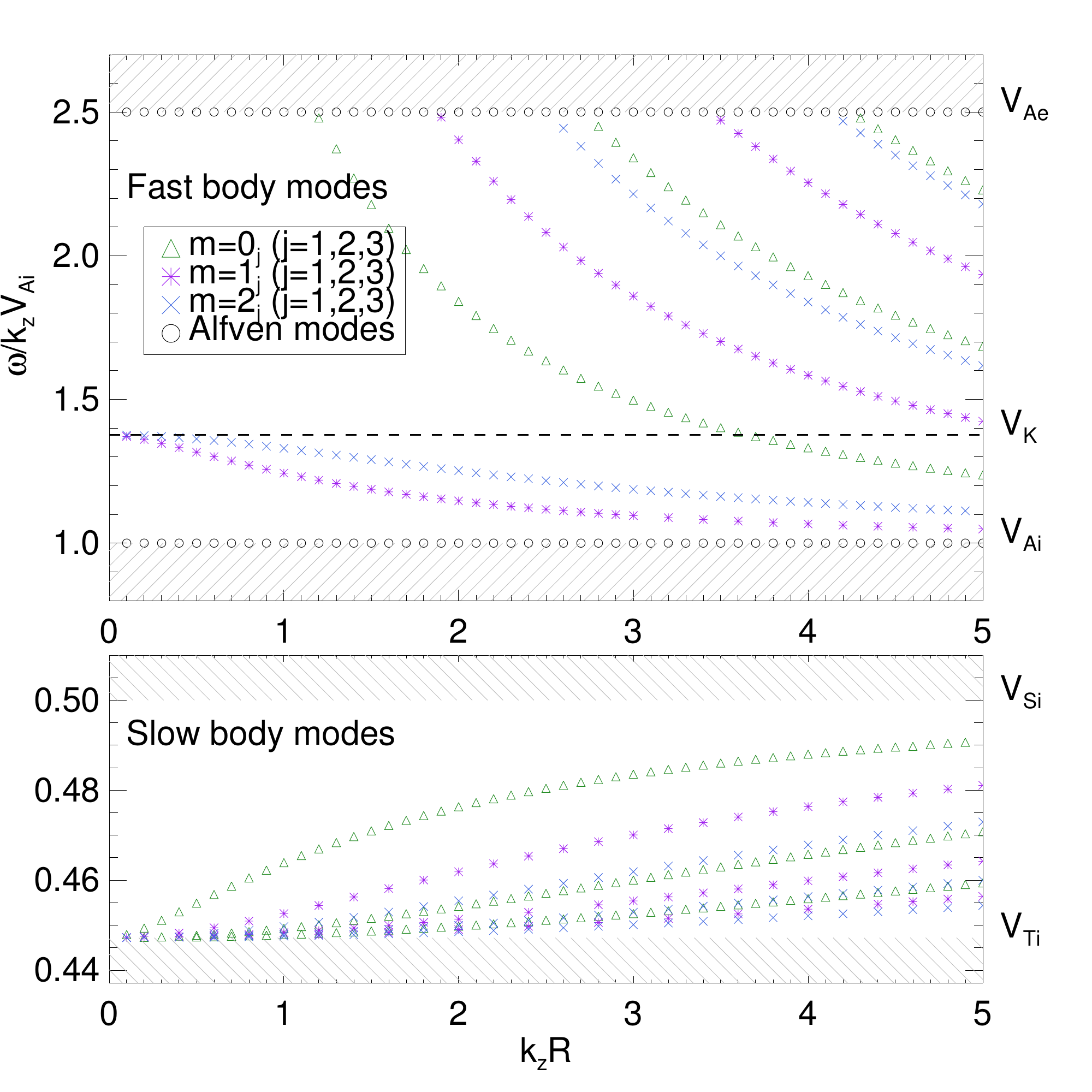}}
	\caption{A graph showing the dispersion curves for the solutions of Equation~\ref{eq:cylinder}. The index $j$ denotes the number of nodes in the radial direction. The wave speeds were taken as $V_\mathrm{Ae}=5V_\mathrm{Si}, V_\mathrm{Ai}=2V_\mathrm{Si}, V_\mathrm{Se}=0.5V_\mathrm{Si}$. The graph clearly shows the different time scales for fast and slow waves.}
	\label{fig:disprel}
\end{figure}
This dispersion relation has infinitely many solutions, leading to a wide variety of possible wave modes in the plasma cylinders \citep{nakariakov2005}. The solutions are shown in Figure~\ref{fig:disprel}. Of particular importance to the QPPs are the axisymmetric modes with $m=0$, of which there are fast sausage modes and slow (sausage) modes. Since these waves are compressive, they have traditionally been considered as possible causes for the intensity modulations during QPPs. The fast modes would be responsible for the short period QPPs (periods up to tens of seconds), while the slow modes would be responsible for the long period QPPs (periods of the order of a minute). The periods of the fast modes are mainly determined by the external Alfv\'en speed \citep[\textit{e.g.}][]{vasheghani2014}, while the period of the slow mode is determined by the internal sound speed, as indicated in Figure~\ref{fig:disprel}. Kink modes (transverse waves with $m=1$) have also sometimes been considered for the interpretation of intermediate periods \citep[\textit{e.g.}][]{kupriyanova2013,kolotkov2015}, because of their intermediate periods determined by the internal Alfv\'en speed.\par
For the generation of QPPs, the reconnection is then assumed to have a constant acceleration rate for energetic particles. The steady outflow to the foot-points is modulated by the oscillations in the flare loop, which allow the periodic precipitation from \textit{e.g.} a magnetic bottle (by the fast sausage mode). Having multiple harmonics in the loop, would then also naturally explain the multiple harmonics seen in the QPPs. In this theory, it is not very obvious how time delays between different wavelength bands may be explained \citep{dolla2012}.\par
Another possibility in this category for the generation of the QPPs is the excitation of propagating or standing slow waves by the evaporation of chromospheric material due to the flare heat deposition there. In numerical experiments, \citet{fang2015} found that these flows result in periodic emission in the foot-points, and can be directly observed in the flare loop body \citep[confirming the interpretation of the observations of][]{kumar2013,kumar2015}. Moreover, such a mechanism may be responsible for the standing slow waves observed in hot flare loops with SUMER \citep{wang2011,yuan2015}. However, it is not yet clear in how far these waves would be observed in whole-Sun observations of flare signals, as QPPs.\par
Even more possibilities are found in the excitation of fast sausage mode wave trains, as proposed by \citet{nakariakov2004} following \citet{roberts1984}. The temporal signature of a wave train is the presence of three phases: the periodic phase, the quasi-periodic phase  when the time profile variates with the highest amplitude and characterized by frequency modulation, and the decay phase. Such wave trains would lead to ``crazy tadpoles'' in wavelet diagrams \citep[see, \textit{e.g.}][]{meszarosova2014}, in which a narrowband tail preceeds a broadband head. The model was put on a firm theoretical footing recently \citep{oliver2015}. The typical wavelet signal serves as a straightforward identification of this mechanism, but so far only a few publications on observations of this type exist \citep{2009AdSpR..43.1479M,2009A&A...502L..13M,2009ApJ...697L.108M, 2013A&A...550A...1K,2013SoPh..283..473M,nistico2014}. \par
An alternative theory for the explanation of flare loop oscillations is the LCR circuit models (LCR stands for the electric circuit equivalent: inductor, capacitor and resistor). These have been reviewed in-depth by \citet{khodachenko2009}. The expected period for the oscillations in that model is between 1 and 30 s, corresponding nicely to some of the QPP periods. However, it is unclear at the moment how and if the LCR pulsations correspond to the straight MHD models of Equation~\ref{eq:cylinder}, and this makes the correspondence with the observations also non-trivial.\par
Previously the thermal overstability and the balooning instability of curved loops were proposed as models for QPP oscillations, but given the lack of recent work, we refer the reader to \citet{nakariakov2009qpp} for more details. The balooning instability was invoked by \citet{zaitsev2013} to explain sub-THz emission in flares.

\subsection{Forward modelling}
Recently, forward modelling (\textit{i.e.} creating artificial observations) has become available \citep[\textit{e.g.}][]{nita2015,vd2016fomo}, thanks to the improved computational power. Especially the {\em GX Simulator} by \citet{nita2015} is directly aimed at the simulation of emission from solar flares, and {\sc FoMo-GS} by \citet{vd2016fomo} computes the gyrosynchrotron emission. Most of the models in Section~\ref{sec:models} only discuss plasma parameters, and do not compute directly the expected emission. \par
It seems intuitive that the compressive fast sausage mode has associated intensity perturbations, and that is why it is popular for explaining the QPPs. However, with forward modelling, it was shown that the intensity variations may be small \citep{gruszecki2012}, and strongly dependent on the spectral line \citep{antolin2013}. Also in the gyrosynchrotron radio emission, an unintuitive dependence on the radio wavelength \citep{reznikova2014,kuznetsov2015} was found, in contrast to earlier analytical predictions \citep{mossessian2012}. \par
Thus, it is clear that the intuitive understanding of the wave modes of the flaring system does not translate directly to the observed emission. Therefore, we would argue that forward modelling of the emission of flare models is essential when trying to identify the correct mechanism for the QPP generation. It is even necessary to go beyond simple analytical approximations \citep[like][]{gruszecki2012,dulk1982} to correctly predict the emission, especially for the optically thick gyrosynchrotron emission \citep{reznikova2014}. The emission near the foot-points is even more difficult to compute, because there non-LTE effects come into play.

\subsection{Seismology}
With seismology, one uses observed wave properties to infer information about the wave medium. Given the high occurrence rate of QPPs in flares \citep{kupriyanova2010}, they have a great potential for seismology for a direct probing of the flare environment. This point is even made more strongly because the flare emission dominates the emission of the whole solar disk, during the brief interval of the flares. Thus, seismology is possible even if no spatial resolution is available. Indeed, the dominating emission from the flare ensures a high spatial resolution, because of the small emitting volume. This is because, in contrast to normal observations (where the instrument resolution is the limiting factor), the inherent high spatial resolution of the flare emission is provided here by the small-scale physics near the X-point and foot-points, which is coming only from a tiny fraction of the coronal volume.  \par
However, before seismology can be performed, it is first necessary to establish the mechanism behind QPP generation. As pointed out before, it could be because of waves in the external medium, periodicities generated in the X-point (by steady processes) or waves in the flaring loop system. This is why there is currently such a great interest in the observational and modelling studies of QPPs: once the generation mechanism of the QPP is established, seismology can reveal new information about the flare surroundings.\par
Some preliminary studies have been undertaken, making an assumption on the physical mechanism behind the observed QPP periods. \citet{vd2011b} assumed that the oscillations were solely taking place in the flaring loop (\textit{i.e.} scenario in Section \ref{sec:postflare}), and made an identification of the modes based on their periodicity. Short periods were taken to be fast sausage modes, while the long periods were assumed to be standing slow modes. While not based on spatial information in this case, these associations were based on the experience gained in spatially resolved observations. With the additional assumption that the wavelengths of both modes are related via a rational number, it was possible to determine the flare plasma $\beta$ to be 0.4. However, the weak point is that the mode identification is only based on temporal scales. \par
More recently, \citet{kupriyanova2013} found QPPs with 3 periods. They assumed that the three periodicities are subsequent harmonics of one oscillation mode and found that the kink mode matches the periods best. Such a multiperiodic event has a lot of potential for seismology: the more periods, the more information can be derived about the medium. A similar approach was taken by \citet{kolotkov2015} who could exclude different wave modes because of unrealistic temperature estimates for slow modes. They found that the long period is likely a kink mode while the short period is a sausage mode. They used the latter to estimate the aspect ratio of the flare loop $L/R$ to be 12.\par
Recently, a new seismology scheme for the inversion of sausage mode periods was developed by \citet{chen2015}. In order to use this, short period QPPs need to be observed of which also the damping time must be confidently measured (and again the assumption that QPPs are caused by sausage modes). Using the period and damping time, one could determine a relation between the Alfv\'en travel time, the width of the inhomogeneous layer and the density contrast of the flaring loop. The measurement of QPP damping times seems to be just around the corner with the current quality of observations.\par
The fact that different wave mode identifications are reported is not surprising and does not contradict the assumption of MHD waves being responsible for the QPPs. Flares are impulsive and localised drivers, and thus contain (mathematically speaking) all wave modes. Therefore, it is to be expected that many wave modes can and will be observed in the flare emission.\par
In most of the seismology models, it is assumed that only one mechanism for the generation of QPPs is at work. However, this may not be true. It could be that the reconnection is modulated by an external wave (Section~\ref{sec:external}), and that the periodic particles are modulated further down by waves in the flare loop (Section~\ref{sec:postflare}), and thus leading to multiple periodicity. In this case, it is very hard to make progress with seismology, and that is why all the current seismological inferences assume one single mechanism for the QPPs.

\section{QPPs in stellar flares}
\begin{figure}
	\centerline{\includegraphics[width=.7\linewidth]{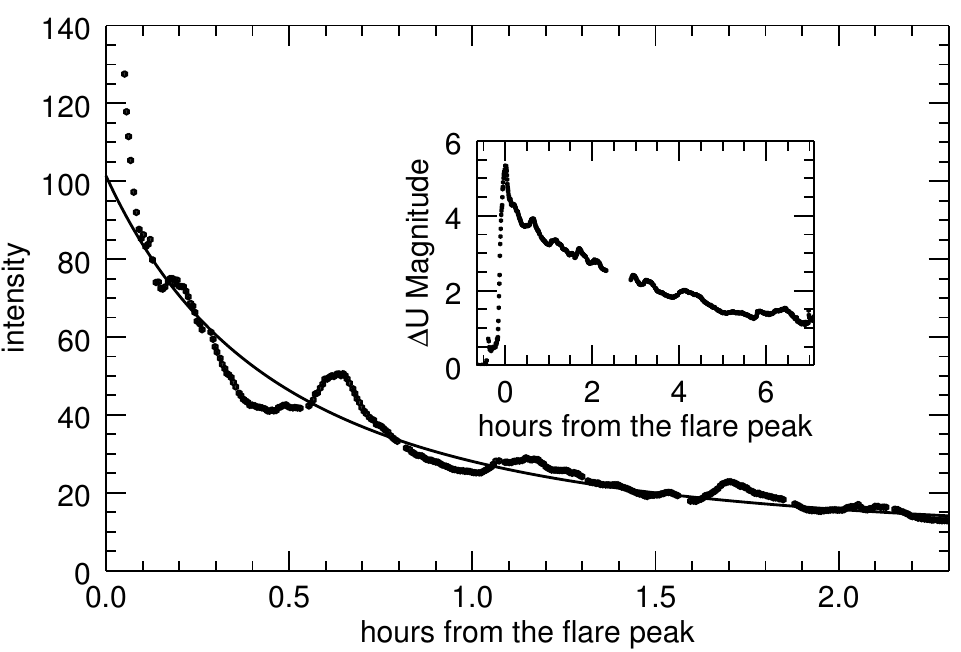}}
	\caption{The emission intensity as a function of time for the megaflare on YZ CMi \citep{kowalski2010}. There is a clear 32 min period QPP present. Figure taken from \citet{anfinogentov2013}.}
	\label{fig:anfinogentov}
\end{figure}

It has long been known that other stars harbour flares as well. This was recently made prominent by the use of Kepler data \citep{walkowicz2011,maehara2012,balona2012,shibayama2013,candelaresi2014}. It is known for a long time that stellar flares show periodic signals \citep{rodono1974}, which can be interpreted as stellar QPPs. Since the turn of the century, we have had clear detection of stellar QPPs by \citet{zhilyaev2000,mathioudakis2003,zaitsev2004,mitra-kraev2005,mathioudakis2006,welsh2006,anfinogentov2013,srivastava2013,balona2015,pugh2015}. The time scale for most of these observations is common over many spectral types and compatible with periodicities in solar flares \citep{maehara2015}. On the other hand, much longer periods (up to a few 1000s, see Figure~\ref{fig:anfinogentov}) are also measured in stellar flare QPPs, but this is perhaps because solar flares are too short (because of their comparatively low amplitude) to sustain (or develop) these periods. \par
Previously prominences were thought to be responsible for a periodic obscuration of the flare emission \citep{houdebine1993}, but we know from solar observations that this is not the case. It is to be expected that the same mechanisms for QPP generation (Section~\ref{sec:models}) are applicable to QPPs in stellar flares. Once the QPP generation mechanism has been determined for solar flares, it is fair game to extend this model to stellar flare QPPs (even though we cannot check, due to the lack of spatial resolution). If MHD waves are found to be responsible for QPPs, this would open up seismology of stellar flares as well. This is an exciting new topic, because it would provide more information on the nature of the flare and, by extension, stellar coronae. This is an important subject to study, because stellar atmospheres, their flares and space weather have an impact on exoplanet habitability.\par
Some first steps in performing stellar QPP seismology have been taken, using the assumption that QPP periods correspond directly to periods of MHD waves. \citet{mathioudakis2003} used white light observations of II Peg to determine the magnetic field strength of 1200 G by interpreting the QPPs as kink modes. Using the same interpretation and technique, \citet{mitra-kraev2005} performed seismology using XMM observations of QPPs on AT Mic, finding a magnetic field of 100 G. \par
\citet{anfinogentov2013} performed an in-depth study of the possible interpretation of the long period (about 2000 s) in the megaflare on YZ CMi (see Figure~\ref{fig:anfinogentov}). It was found that the long period could be due to a standing slow wave in a flaring loop with a length of 200 Mm, compatible with earlier modelling estimates. \par
\citet{srivastava2013} observed two periods ($\sim$1300 s and $\sim$690 s) in a QPP on Proxima Centauri. They found that the period ratio is 1.83. Assuming the model of standing kink waves in the flaring loop, they determined the density stratification to be 23 Mm in the stellar corona.\par
The seismology result of \citet{zaitsev2004} used radio observations of AD Leo to study QPPs, and found that the LCR model could explain the observed short periodicities. They estimated the loop length to be on the order of 400~Mm, which is comparable to the M-dwarf stellar radius.\par

\section{Conclusions} 
      \label{sec:conclusion} 

In this article, we have reviewed the observational and theoretical results on quasi-periodic pulsations (a.k.a. QPPs). These QPPs are periodic intensity modulations in the flare emission (across all wavelength bands). In this paper, we have focussed on reporting on the works since the publication of the review by \citet{nakariakov2009qpp}. \par
Despite the many observational and theoretical advances in the last years, it has not been possible to determine what physical mechanism is responsible for causing the QPPs. There are three main lines of thought: (i) the periods could be created by external modulation of the reconnection (Section~\ref{sec:external}), (ii) the periods could be inherent to the reconnection process (Section~\ref{sec:flare}), or (iii) the periods are modulations of the flare particles by waves in the flaring loop (Section~\ref{sec:postflare}). We believe that the determination of the correct mechanism (at least for the long-period QPPs) will take place in the next few years, by comparing spatial information from the EUV to the temporal emission, where recent results have been achieved. It could be that the ALMA instrument will play a role in this \citep{wedemeyer2015}. \par
We stress that QPPs carry a lot of potential for further research. Once the correct mechanism has been identified (and it is related to MHD waves), it is possible to use them for seismology: by comparing the QPP periods to models, we can determine physical quantities near the flare site. \par
An interpretation in terms of MHD waves in the flaring loops (Section~\ref{sec:postflare}) is currently the most popular, because it explains most of the observational features (such as multi-periodicity). Making this assumption, it is possible to make progress in measuring flare parameters seismologically. Most of the solar and stellar QPP seismology models are currently based on crude models. However, it is a promising start, and more advanced models can lead to better understanding of the flare surrounding. In any case, it is important that flare models naturally incorporate QPPs, because they seem to occur in most flares.\par
In our view, progress in the understanding of QPPs can be mainly made in two ways:
\begin{enumerate}
	\item Observationally identifying the source location for the QPPs.
	\item Constructing realistic models for flares and their QPPs, which include forward modelling. 
\end{enumerate}

\begin{acks}
 Inspiration for this review was obtained during ISSI and ISSI-BJ workshops. TVD thanks the Odysseus type II funding (FWO-Vlaanderen), IAP P7/08 CHARM (Belspo), GOA-2015-014 (KU~Leuven). EK is a beneficiary of a mobility grant from Belspo.
 
 Disclosure of Potential Conflicts of Interest: The authors declare that they have no conflicts of interest.
\end{acks}

\bibliographystyle{spr-mp-sola}
\bibliography{qpprefs,qpprefs_ek}  

\end{article} 

\end{document}